\documentclass[twocolumn,nofootinbib,jcp,preprintnumbers,amsmath,amssymb]{revtex4-1}
\usepackage{natbib,hyperref}
\usepackage{amsmath}
\usepackage{graphicx}
\usepackage{dcolumn}
\usepackage{bm}
\usepackage{epsfig,color,xspace,multirow,xr,bbold}
\usepackage[all]{xy}
\usepackage{setspace}
\usepackage{url}
\usepackage[colorinlistoftodos]{todonotes}
\usepackage{threeparttable}

\usepackage{natbib,hyperref}
\usepackage{float}
\usepackage{placeins}
\usepackage[utf8]{inputenc}

\begin{document}

\title{
Charge-Transfer Selectivity and Quantum Interference in Real-Time \\
Electron Dynamics: Gaining Insights from \\
Time-Dependent Configuration Interaction Simulations
}
\date{\today}     
 
\author{Raghunathan Ramakrishnan$^{1}$}
\email{ramakrishnan@tifrh.res.in}

\affiliation{$^1$Tata Institute of Fundamental Research, Centre for Interdisciplinary Sciences, Hyderabad 500107, India}


\begin{abstract}
Many-electron wavepacket dynamics based on time-dependent configuration interaction (TDCI) is a numerically rigorous approach to quantitatively model electron-transfer across molecular junctions.  TDCI simulations of cyanobenzene thiolates---para- and meta-linked to an acceptor gold atom---show donor states \emph{conjugating} with the benzene $\pi$-network to allow better through-molecule electron migration in the para isomer compared to the meta counterpart. For dynamics involving \emph{non-conjugating} states, we find electron-injection to stem exclusively from distance-dependent non-resonant quantum mechanical tunneling, in which case the meta isomer exhibits better dynamics.  Computed trend in donor-to-acceptor net-electron transfer through differently linked azulene bridges agrees with the trend seen in low-bias conductivity measurements. Disruption of $\pi$-conjugation has been shown to be the cause of diminished electron-injection through 1,3-azulene, a pathological case for graph-based diagnosis of destructive quantum interference.  Furthermore, we demonstrate quantum interference of many-electron wavefunctions to drive para- vs. meta- selectivity in the coherent evolution of superposed $\pi$(CN)- and $\sigma$(NC-C)-type wavepackets.  Analyses reveal that in the para-linked benzene, $\sigma$ and $\pi$ MOs localized at the donor terminal are \emph{in-phase} leading to constructive interference of electron density distribution while phase-flip of one of the MOs in the meta isomer results in destructive interference. These findings suggest that \emph{a priori} detection of orbital phase-flip and quantum coherence conditions can aid in molecular device design strategies.
\end{abstract}

\maketitle
\section{Introduction}
First-principles understanding of why, how and how much electric current flows 
across a given molecule holds the key to unlock
challenges in designing electronic circuits of sub-nanometer dimensions
with atomistic precision\cite{aviram1974molecular,reed2000computing,heath2003molecular,nitzan2003electron,joachim2000electronics,su2016chemical,weber2002electronic,aradhya2012dissecting}. 
Pioneering efforts in the design of
scanning tunneling microscope-based break junction (STM-BJ) experiments
have made it feasible to accurately determine the
{\it through-molecule} conductance, $G=I/V$, of a device 
at vanishing bias-voltage at which the molecular electronic 
structure is least perturbed\cite{venkataraman2006dependence}. 
A histogram of $G$ is made by recurrently forming and breaking the 
contacts between molecules and the STM tip
where spikes for values less than the quantum of conductance 
$G_0=2e^2/h\thickapprox7.75\times10^{-5}$\,S indicate flow of 
current through single molecule junctions\cite{xu2003measurement}. 
As for first-principles modeling of quantum conductance, the standard
approach is the Landauer formalism for coherent 
transport---valid for short  junctions at low 
temperatures---wherein electrons flow across molecules through conduction channels that are related to the molecular orbitals (MOs)\cite{cuniberti2006introducing}. At zero bias voltage, conductance  is calculated as $G(E,V) = G_0 \sum_{i,j} T_{i,j}(E,V)$, where $T_{i,j}$ is the probability that a charge carrier coming from a terminal in transverse channel $i$
will be transmitted to another terminal in channel $j$. This 
formalism based on the nonequilibrium
Green’s function (NEGF) method\cite{datta2005quantum} 
has found wide applicability when used with MOs 
modeled at various quantum chemistry 
levels of theory ranging from the empirical H\"uckel MO (HMO)
model Hamiltonian\cite{pedersen2015illusory} to the 
Kohn--Sham density-functional theory (KS-DFT)\cite{brandbyge2002density}. 
For prototypical systems, both NEGF-HMO and NEGF-DFT
methods have shown to give qualitatively similar transmission
spectra; while the former accounts only for $\pi$-tunneling, 
the latter approach not only captures
tunneling via all the MOs but also provides 
quantitatively accurate treatment of electronic interaction\cite{koga2012orbital}.


Quantum interference (QI) is an experimentally quantifiable  effect
stemming from the phase differences of the current flowing through  multiple pathways within a molecular junction\cite{buttiker1986quantum,sautet1988electronic,ke2008quantum}. 
Interest in QI had its beginnings from observations on a mesoscale metal 
ring ($\thickapprox$ $\mu m$ in diameter) where the resistance 
as a function of applied magnetic field displayed oscillations characteristic of 
the Aharonov--Bohm effect\cite{datta1997electronic}. One of the prime physical 
factors that destroy the QI effects is inelastic scattering during conduction. 
Minimizing such scattering effects requires the {\it loop} dimension 
to be of the size of the benzene molecule\cite{nitzan2003electron}.
Subsequent investigations in this direction have largely been motivated by the 
experimental demonstration of meta-vs.-para ({\it m}-vs.-{\it p}) selectivity 
in conduction across benzene; for instance, Mayor {\it et al.} using the STM-BJ 
technique demonstrated current flowing through two {\it m}-linked benzene rings 
to be two orders of magnitude smaller than that flowing through the {\it p}-linked analogue\cite{mayor2003electric}.
Similar conclusions have been drawn also in somewhat more recent experiments on
coupled benzene rings\cite{arroyo2013signatures}.
More recent experimental efforts have even established such 
subtle correlations like
the effects bond topology and electronegativity of atomic sites can have 
on the degree and the location of QI features in 
molecular wires\cite{zhang2018controlling}. 
It is suggested that a combination of
stimuli-response and QI can be an
efficient strategy to enhance isomer recognition and
conductance switching in single-molecule junctions\cite{zhang2018distinguishing}.
While conjugation has widely been considered as a main tool to control QI, 
recent synthetic efforts have shown that the effect can be manipulated through
chemical modification of the molecular wire\cite{naghibi2019synthetic}. 
Exploiting QI for practical
purposes requires that decoherence effects are minimal; 
it is an experimentally established fact 
that with increase in temperature, 
destructive interference effects are lost resulting in 
enhanced conductivity\cite{ballmann2012experimental}. 

\begin{figure}[htp!] 
\centering          
\includegraphics[width=8.5cm, angle=0.0]{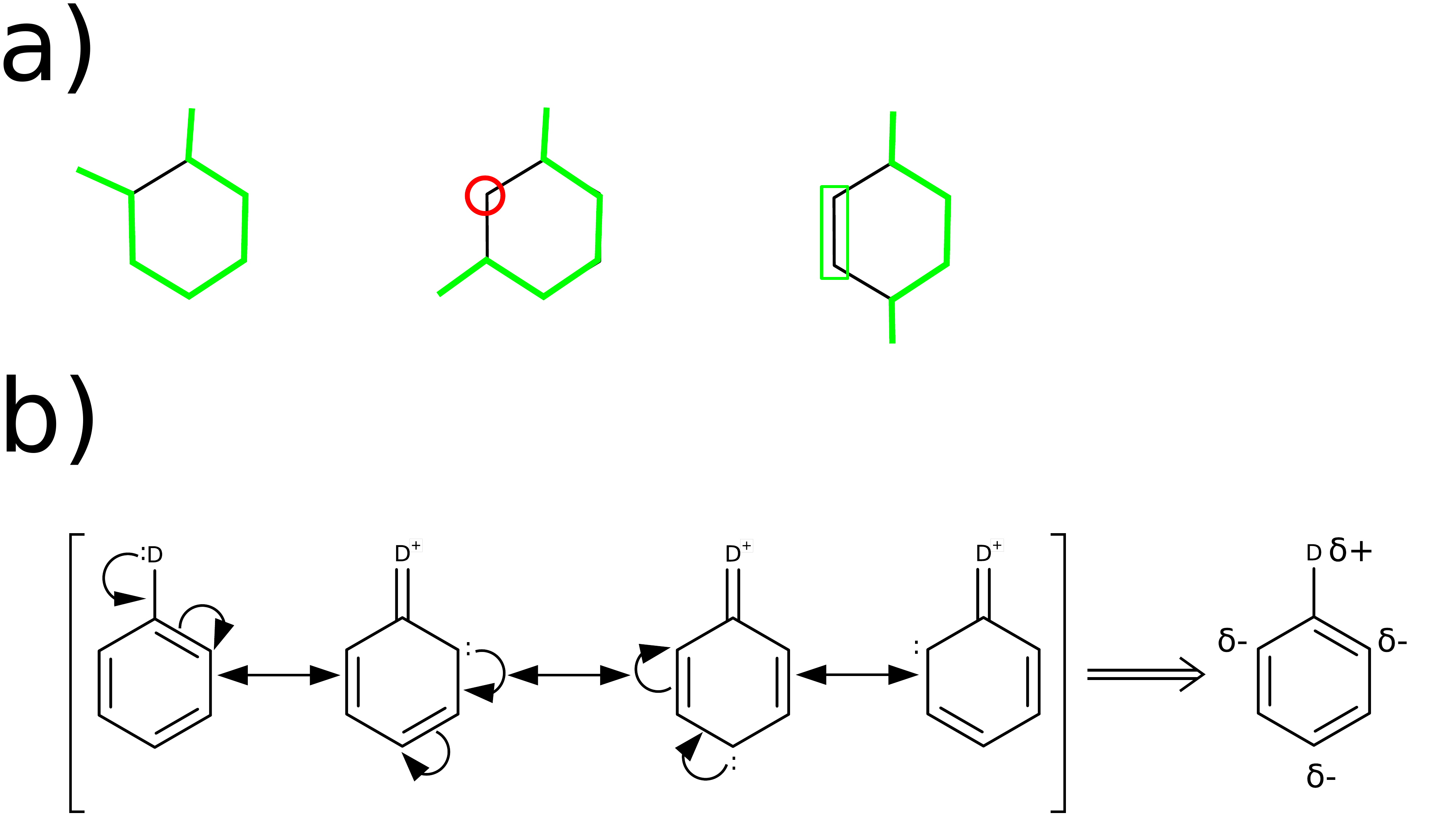}
\caption{
Prediction of QI features based
on molecular graphs and electronic structure. 
a) Connectivity rules
based on the longest continuous paths in 
{\it o-}/{\it m-}/{\it p-}linked benzenes\cite{markussen2010relation}; 
an isolated atomic center not located 
on the path is marked with a red circle in the $m$-linked benzene.
b) Curly arrow rules inspired by aromatic resonance stabilization effects\cite{stuyver2015back,hosoya2015cross}.}
\label{fig:QIscheme}
\end{figure} 
From a theoretical stand point, a number of studies have established 
qualitative relationships between observed/predicted QI trends with conjugation patterns 
in hydrocarbons---destructive QI in cross-conjugated molecules and constructive QI in linearly conjugated ones\cite{reuter2014communication,tsuji2018quantum,nozaki2017molecular}. In particular, Markussen {\it et al.} have presented a set of graphical rules to predict if a molecular 
structure can lead to 
QI or not\cite{markussen2010relation}. Accordingly, a molecule will feature destructive QI if the longest continuous
path that can be drawn across it connecting both terminals leaves at 
least one  atomic site unconnected ({\it i.e.} unpaired) and without a nearest neighbor. 
Using this rule, it is straightforward to see
why an {\it m}-linked benzene junction with a single unpaired site will suffer 
from destructive QI amounting to diminished conductivity (see Fig.~\ref{fig:QIscheme}a).
Others have presented a selection rule for QI based on curly arrow
diagrams that are traditionally used to diagnose
 resonance stabilization patterns in $\pi$-electron
 conjugated systems\cite{stuyver2015back,hosoya2015cross}.
 Fermi-level destructive QI is noted in a conjugated molecule when curly arrows 
 cannot displace an electron pair from the donor end to at least one of the
other sites; Fig.~\ref{fig:QIscheme}b showcases how in benzene an electron 
pair cannot be displaced from the donor end to the {\it m} positions. 
In this context, it may be worthwhile to note that in a 
 theoretical study using density-matrix propagation based on the tight-binding $\pi$ 
Hamiltonian, blocking one of the paths in a {\it p}-linked benzene has resulted 
in essentially no change in the dynamics of current flow
 indicating the lack of interference in the {\it p}-isomer
as far as only the $\pi$-channels are concerned\cite{chen2014interference}. 
The same study showed that an {\it m}-linked isomer, 
initially showing poor current flow because of phase coherence 
(or destructive QI) between the current
 flowing through two paths, exhibiting much improved dynamics and better 
 rates when one of the paths is blocked. 

Interestingly, in contrast to the trends noted for molecular conductivity, 
Gorczak {\it et al.} observed in photoinduced charge transfer (CT) measurements 
of donor-bridge-acceptor (D-B-A) systems
faster hole-transfer timescales in cross-conjugated junctions compared 
to linearly conjugated ones\cite{gorczak2015charge}.
Furthermore, this study has reported D-B-A hole transfer via an {\it m}-linked 
biphenyl bridge---with a shorter D-A throughspace distance---to be faster than via 
a {\it p}-linked isomer. The through-{\it m} channel
also benefits from contributions from the $\sigma$-type-MOs 
leading to faster CT timescales at least 
in shorter molecular junctions\cite{gorczak2015charge,borges2016probing}. 
More recent experimental studies have stressed 
that for a successful rational design of molecular
junctions, an understanding of QI effects in $\sigma$-channels is as important as those
in $\pi$-channels\cite{garner2018comprehensive}.
The close relation between molecular conduction and D-to-A 
electron-transfer properties has been discussed 
by others\cite{nitzan2001relationship}. Meanwhile, somewhat different CT
trends have been noted in longer bridge molecules such as  
cross-conjugated xanthone which shows 30 times slower charge-injection dynamics
compared to the linearly-conjugated molecule trans-stilbene. In the former case, 
it has been argued that cross-conjugation strongly decreases the 
$\pi$ orbital contribution to D-A electronic coupling so that electron transfer 
most likely uses the bridge $\sigma$ system as its primary pathway\cite{ricks2010controlling}.

It is the purpose of this article to complement 
continuously evolving chemical intuitions about transport selectivity 
across isomeric molecular junctions with electron dynamics modeling accounting for
many-body coherence and electron correlation effects.
To this end, the formally exact formalism of time-dependent
configuration interaction (TDCI)\cite{klamroth2009ultrafast,krause2005time} has been employed to ${\it see}$ electron dynamics in cyanobenzene and in $m$/$p$-linked benzonitrile thiolate (CN-C$_6$H$_4$-S-)
molecules bonded to a gold (Au) atom serving as the
acceptor terminal. To showcase the applicability of the TDCI methodology towards understanding
electron transfer  selectivity in non-alternant hydrocarbons, we have studied 
cyanoazulene thiolate molecules (CN-C$_{10}$H$_6$-S-) linked to an Au atom through 
four different substitution patterns.

\section{Methods}
\subsection{Time-dependent configuration interaction}
Within the scope and restrictions of the Born-Oppenheimer approximation,
any electronic property of a molecule, with a corresponding quantum mechanical operator 
$\hat{P}$, can be calculated as a function of time
once we have the time-dependent (TD) wave function $\Psi_{\rm e}({\bf r},t)$ obeying the 
time-dependent Schr\"odinger equation (TDSE)
\begin{eqnarray}
i\frac{\partial}{\partial t} \dot{\Psi}_{\rm e}({\bf r},t)=\hat{H}_{\rm e}\Psi_{\rm e}({\bf r},t),
\end{eqnarray}
where $\Psi_{\rm e}({\bf r},t)$ and $\hat{H}_{\rm e}$
are the electronic wavefunction and the electronic Hamiltonian, respectively.
The TDCI approach is formally exact as long as the
CI wavefunction is expanded with all possible configuration state functions (CSFs). This incurs very heavy computational requirements and renders all but molecules of the size of water tractable. In this study, we have truncated the CI expansion to up to 
singles and doubles substitution ({\it i.e.} CISD): 
\begin{eqnarray}
| \Psi \rangle = 
c_0| \Psi_0^{\rm HF}\rangle + 
\sum_{a,r} c_{a}^{r} | \Psi_{a}^{r}\rangle + 
\sum_{a<b;r<s} c_{a,b}^{r,s} | \Psi_{a,b}^{r,s}\rangle,
\end{eqnarray}
where $|\Psi_0^{\rm HF}\rangle$ is the Hartree--Fock ground state; 
$| \Psi_{a}^{r}\rangle$ and $| \Psi_{a,b}^{r,s}\rangle$ denote singly and doubly substituted Slater determinants, respectively,
with $a,b$ going over the indices of occupied spin orbitals while $r,s$ are indices of the unoccupied spin orbitals. The total number of Slater
determinants entering the expansion of the CISD wavefunction
scales as ${\mathcal O}(N_o N_v)$ and ${\mathcal O}(N_o^2 N_v^2)$ ($N_o$ and $N_v$ are number of occupied and virtual MOs) 
for singles and doubles substitution, respectively. Since the electronic states studied in this work are of singlet-spin type, a more efficient approach is to represent $\Psi({\bf r},t)$ in the 
variational space spanned by singlet spin-adapted CSFs\cite{szabo1996modern}:
\begin{eqnarray}
 |^1 \Psi_a^r \rangle & = & \left[ | \Psi_{\bar a}^{\bar r} \rangle + | \Psi_a^r \rangle  \right] / \sqrt{2} \nonumber \\
 |^1 \Psi_{aa}^{rr} \rangle & = &  | \Psi_{a\bar a}^{r\bar r} \rangle \nonumber  \\
 |^1 \Psi_{aa}^{rs} \rangle & = & \left[ | \Psi_{a \bar a}^{r\bar s} \rangle + | \Psi_{a\bar a}^{s\bar r} \rangle  \right] / \sqrt{2} \nonumber \\
 |^1 \Psi_{ab}^{rr} \rangle & = & \left[ | \Psi_{\bar a b}^{\bar r r} \rangle + | \Psi_{a\bar b}^{r\bar r} \rangle  \right] / \sqrt{2} \nonumber \\
 |^A \Psi_{ab}^{rs} \rangle & = & \left[ 
 2| \Psi_{\bar a b}^{\bar r s} \rangle
 +2| \Psi_{\bar a \bar b}^{\bar r \bar s} \rangle
 -| \Psi_{\bar a b}^{\bar s r} \rangle
 +| \Psi_{\bar a b}^{\bar r s} \rangle +| \Psi_{a \bar b}^{r \bar s}  \rangle \right.  - \nonumber \\
 & & \left. | \Psi_{a \bar b}^{s \bar r} \rangle \right] / \sqrt{12} \nonumber \\
 |^B \Psi_{ab}^{rs} \rangle & = & \left[ 
 | \Psi_{\bar a b}^{\bar s r} \rangle
 +| \Psi_{\bar a b}^{\bar r s} \rangle
 +| \Psi_{a \bar b}^{r \bar s} \rangle
 +| \Psi_{a \bar b}^{s \bar r} \rangle
  \right] / 2.
\end{eqnarray}
The notation conveys that spin orbital indices with an 
overline denote beta-spin electrons and those without, alpha-spin electrons.
The TDSE is solved as an initial value problem, where 
qualitative trends in electron dynamics depend on the choice of the
initial state.  

For the dynamics to result in an efficient CT process,
the initial state, $| \Psi(0) \rangle$ must satisfy the following formal
criteria: Firstly, since the process being simulated is a field-free evolution with
conserved total energy, the initial
state must be non-stationary, {\it i.e.}, formally a wavepacket that is a 
linear superposition of the electronic energy eigenstates. 
Secondly, the real-space picture of the initial state must be such that 
in the neighborhood of the wavepacket's energy, 
there is a net difference in the density-of-states (DOS) between geometric 
ends of the molecule. The donor terminal is typically that with an excess density
of occupied MOs while that with excess unoccupied MOs mark the acceptor terminal.   
Experimentally, such an initial electronic configuration
can be created in core-hole-clock spectroscopy, where typically,
an electron from a main-group atom such as nitrogen is excited to the $\pi^*$
MO localized on the CN 
terminal\cite{hamoudi2011orbital,blobner2012orbital,fohlisch2005direct}.  
Finite size of the acceptor terminal or finiteness of the states that are localized on the 
acceptor terminal results in a situation where residual 
electron density shuttles back and forth between the donor
and acceptor ends as noted before in Li-terminated D-B-A molecules\cite{ramakrishnan2013electron}. In order to stabilize the CT dynamics, electron trapping is essential. This requires 
that the acceptor end is made either of a single transition metal atom with 
several unoccupied orbitals or a metal cluster 
with large number of vacant MOs. Of utmost importance
is also the fact that the bridge region of the molecule must have a large density of vacant MOs 
serving as the conduction band. Finally, it is important to note that field-free D-B-A CT dynamics is symmetry-controlled\cite{blobner2012orbital}.

\begin{figure*}[htp!] 
\centering          
\includegraphics[width=16cm, angle=0.0]{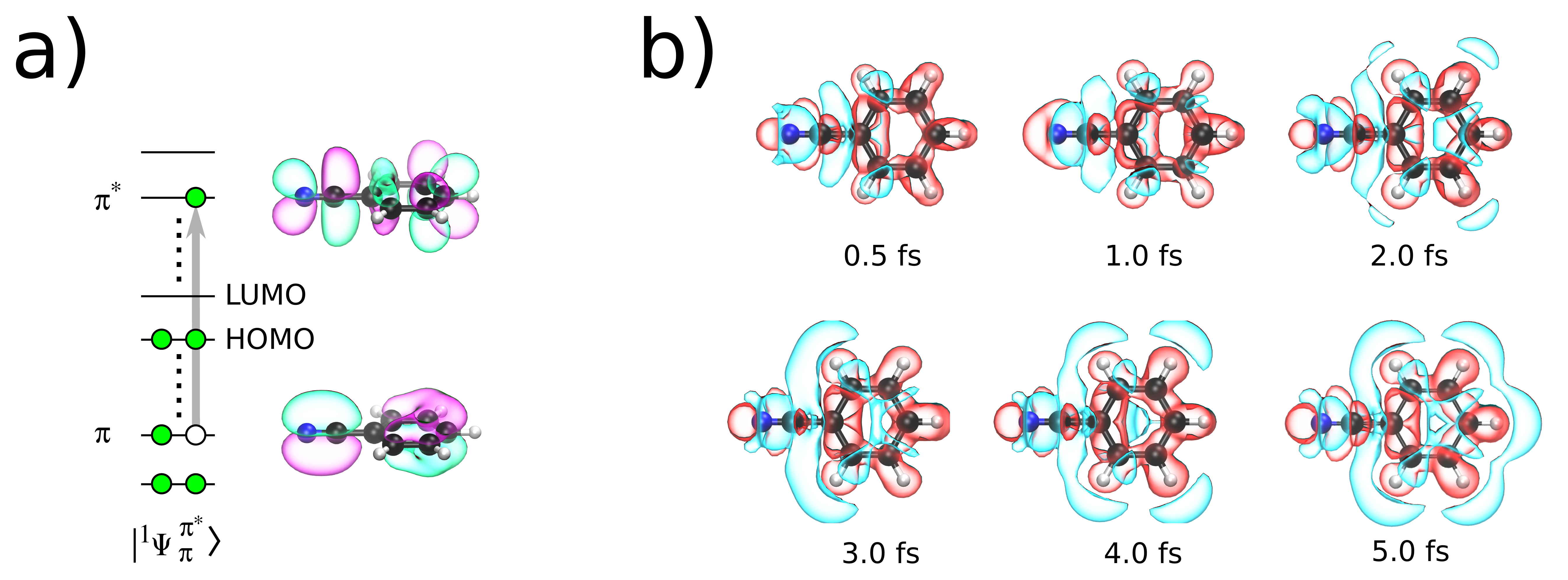}
\caption{
TDCI wavepacket dynamics of cyanobenzene:
a) Creation of a non-stationary electronic state in cyanobenzene by an implicit 
$\pi \rightarrow \pi^*$ single-excitation; plots of the corresponding MOs are shown, 
b) Electron density difference, 
$\Delta \rho({\bf r}, t) = \rho({\bf r}, t) - \rho({\bf r}, 0)$, 
is shown for the first few fs of electron dynamics;
blue and red color indicate gain and depletion, respectively.}
\label{fig:fig02}
\end{figure*}

The TD partial charge $q(t)$ on an atom $A$ can be 
computed by summing over all atomic orbitals (AOs) 
$\mu$ centered on that atom according to L\"owdin's formula
\begin{eqnarray}
q_{A} (t) = Z_A - \sum_{\mu \in A} \left[ \mathbf{S}^{1/2} \mathbf{P}(t) \mathbf{S}^{1/2} \right]_{\mu \mu},
\end{eqnarray}
where $\mathbf{S}$ is the overlap matrix in the AO representation, 
and $\mathbf{P}(t)$ is the TD reduced charge-density bond-order (CDBO) matrix:
\begin{eqnarray}
P_{\mu,\nu}(t)&=&
\langle \mu | {\mathrm {Tr}}_{2,\ldots,N} |\Psi({\bf r},t) \rangle \langle \Psi({\bf r},t)| \nu \rangle.
\end{eqnarray}
These matrix elements can be computed for a many-body wavefunction 
by applying the Slater--Condon rules\cite{szabo1996modern,ulusoy2011correlated}. 
With the reduced CDBO matrix, one can plot the three-dimensional electron
density as a function of time through
\begin{eqnarray}
    \rho({\bf r},t) & = & \sum_{\mu, \nu} \phi_{\mu}({\bf r}) P_{\mu,\nu}(t) \phi_{\nu}({\bf r}), 
\end{eqnarray}
where $\phi_{\mu}({\bf r})$ is an AO.

\subsection{Computational details}
Minimum energy equilibrium structures of 
cyanobenzene (C$_6$H$_5$CN),
{\it p}/{\it m}-linked CN-C$_6$H$_4$-S-Au isomers,
and CN-C$_{10}$H$_6$-S-Au isomers
 were optimized at the 
KS-DFT level, PBE0\cite{adamo1999toward}, using the quantum chemistry package 
NWCHEM~(version 6.6)\cite{valiev2010nwchem}.
The split-valence basis set def2-SV(P)\cite{weigend2005balanced} containing a polarization function was employed for all atoms. 
For Au, an effective core potential, ECP60MWB\cite{andrae1990energy}, was used to replace the 60 core electrons while the remaining 19 
electrons---accounting for the 
valence configuration 5$s^2$5$p^6$5$d^{10}$6$s^1$---were treated explicitly with the 
aforementioned basis set. 
All TDCI calculations have been performed using locally developed codes\cite{ramakrishnan2013electron,ramakrishnan2015charge}. 
The present implementation depends on one- and two-electron molecular integrals along with
Hartree--Fock (HF) MOs computed using NWCHEM. Furthermore, all 
CI calculations have been performed in the framework of spin-adapted CSFs with the CI wavefunction expansion truncated by including up to double substitutions ({\it i.e.} CISD).
Along with the HF Slater determinant $|\Psi_0\rangle$, we included 
all possible singly-substituted CSFs, 
$|^1\Psi_a^r\rangle$, and we restricted the active space for doubly-substituted CSFs,
$|^1\Psi_{aa}^{rr}\rangle$,
$|^1\Psi_{aa}^{rs}\rangle$, 
$|^1\Psi_{ab}^{rr}\rangle$,
$|^A\Psi_{ab}^{rs}\rangle$,
$|^B\Psi_{ab}^{rs}\rangle$, to ($N=20$,~$M=60$), where $N$ and $M$ are the number of valence electrons and number of spin orbitals, respectively. For cyanobenzene, {\it m}-/{\it p}-CN-C$_6$H$_4$-S-Au isomers,
and  CN-C$_{10}$H$_6$-S-Au isomers
the resulting restricted-active-space-CI (RASCI)\cite{hochstuhl2012time}
wavefunctions contain
22,666, 25,337 and 29,563 CSFs (62,231, 67,573  and 76,025 
Slater determinants), respectively.
All TD electronic wavepacket propagations were peformed within the 
fixed-nuclei approximation, which is valid for ultrashort time scales\cite{ulusoy2012remarks}. We solved the TDCI equations\cite{krause2005time} using the fourth-order Runge--Kutta method (RK4) with a finite time-step of $\Delta t = 0.001/4 \pi {\mathrm c}{\mathrm {Ry}}=0.024$ attoseconds (as), $1$ as $=10^{-18}$ s.

\section{Results and Discussions}
\subsection{Ultrafast electron dynamics in cyanobenzene}
As a prototype model to illustrate TDCI-based electron dynamics, we begin with 
the simulation of a field-free time-evolution of an electronic wavepacket in the planar
molecule C$_6$H$_5$CN.  
Preparation of an initial state for the TDCI dynamics is sketched
in Fig.~\ref{fig:fig02}a.
Since we would like to understand the participation of 
the out-of-plane $\pi$-type MOs on the benzene fragment---to quantify the 
relative role of the fragment MOs on the {\it o}-/{\it m}-/{\it p}-C atoms---we have chosen
a CSF corresponding to the $\pi\rightarrow\pi^*$ excitation, where an electron from 
the occupied $\pi$ MO is excited to the unoccupied $\pi^*$ MO. The symmetry of 
this CSF belongs to the $a^{\prime\prime}$ irreducible representation of the $C_s$
point group. Furthermore, $| \Psi(0) \rangle$ features an electronic
arrangement that is suitable for CT dynamics satisfying the criteria 
discussed above. Specifically, at $E=\langle \Psi(0)| \hat{H} | \Psi(0) \rangle$
the projected density of states (PDOS)---corresponding to occupied MOs---is larger on the  
CN fragment (the donor terminal) compared to the PDOS on the benzene fragment 
(acceptor terminal).

\begin{figure*}[hpt!] 
\centering                                                                 \includegraphics[width=16.0cm, angle=0.0]{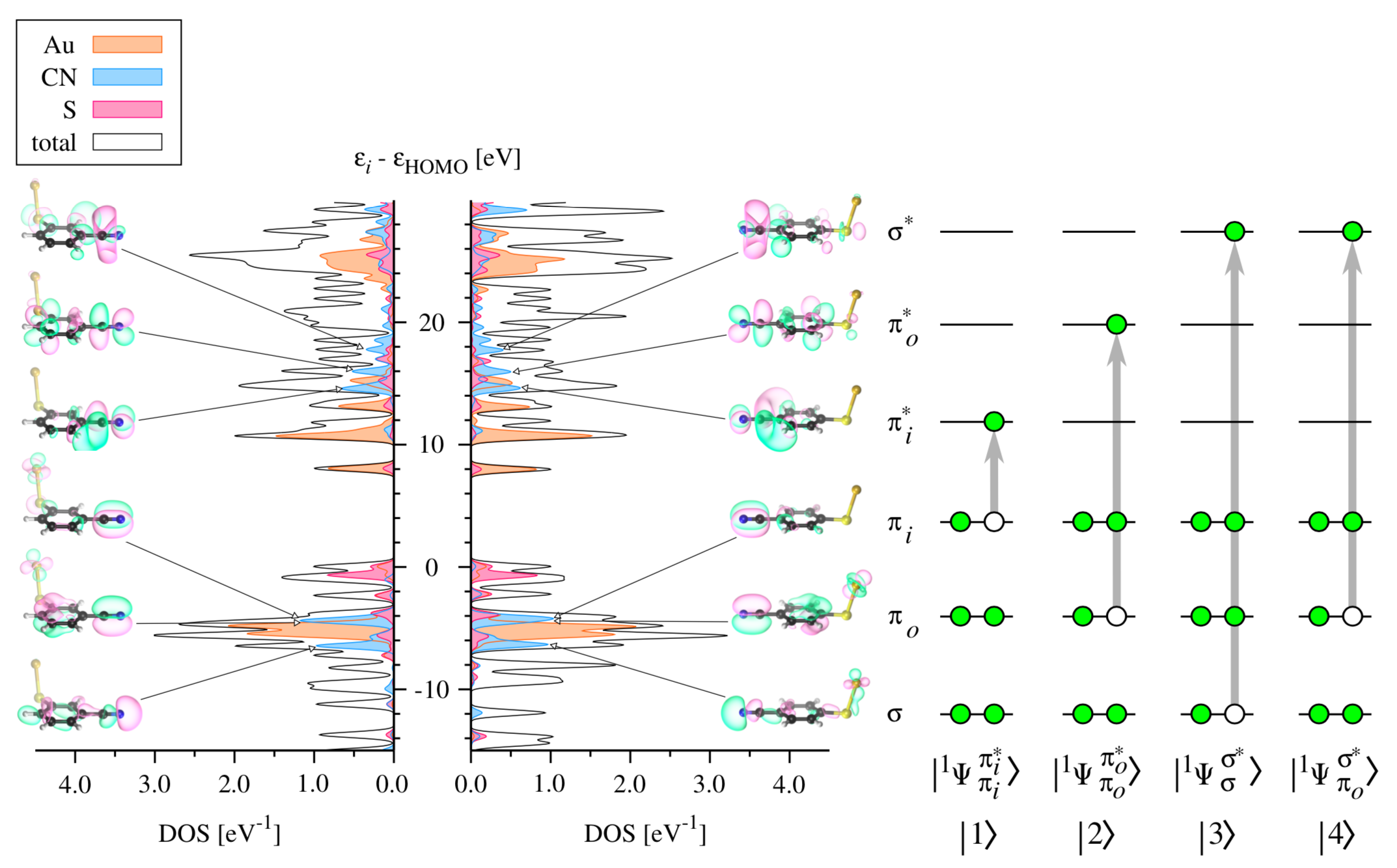}                                       
\caption{
Electronic structure of {\it m} and {\it p} isomers of 
gold-benzonitrile thiolate (CN-C$_6$H$_4$-S-Au). 
For both molecules, projected density of states (PDOS) is plotted with 
insets showing selected MOs.
Also given are the definitions of various singly substituted CSFs that are 
used as initial states in TDCI electron dynamics simulations.  
}
\label{fig:fig03}
\end{figure*} 

Typically, a real-space picture of D-B-A electron dynamics  
show electron-injection from the donor terminal to the acceptor end 
through regions of space localized on the
molecular framework suggesting through-bond CT with sufficient directionality. 
To shed more light on this process, we have plotted the
time-dependent electron density, $\rho({\bf r},t)$, for the 
first few fs of time-evolution. The dynamics proceeds with a very rapid
sub-fs event of re-filling of vacant MOs localized at the CN terminal by 
electron density migration from the benzene fragment. Such a rapid dynamical feature 
is characteristic of strongly coupled donor and acceptor states arising 
from good spatial overlap. Further, the energy gap between $\pi$ and $\pi^*$
MOs considered for the excitation is $15.6$ eV, which corresponds to
sub-fs oscillations according to $(\Delta t~{\rm 0.13~fs})=2.07/ (\Delta E~{\rm 15.6~eV})$;
this process is reminiscent of the sub-fs dynamics arising from the ionization of
a core electron in nitrosobenzene described by Kuleff {\it et al.}\cite{kuleff2016core}.
Focussing on  Fig.~\ref{fig:fig02}b, 
by $t=2$ fs, we note the electron density from the 
CN terminal to get injected into the {\it o}- and {\it m}- C sites of benzene.
Following brief oscillatory dynamics, by  $t=5$ fs, we also note the 
{\it p} sites of benzene to be populated. Beyond 5 fs, as is common in a system with
finite DOS localized at the acceptor terminal, the dynamics show recurrences with partial-revival lifetimes of the order of a few fs. Overall, the TDCI electron dynamics of cyanobenzene
do not indicate {\it m}-vs-{\it p} selectivity in the population transfer from the 
CN group to the C sites of the benzene. The CT timescales to populate various sites
is strongly dependent on the distance between these sites to the CN group. 
\begin{figure*}[hpt!] 
\centering                                                                 
\includegraphics[width=16.0cm, angle=0.0]{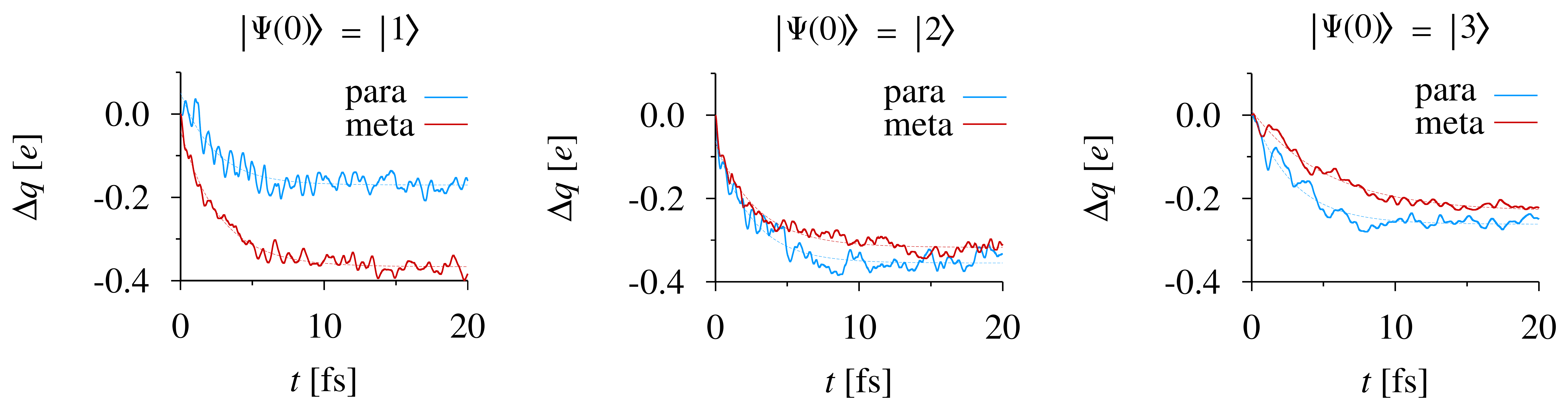}
\caption{
Time evolution of the partial charge, $q$, on the Au atom 
in {\it m}- and {\it p}-linked CN-C$_6$H$_4$-S-Au
for three choices of the
initial state, $|\Psi(0)\rangle$ (see Fig.~\ref{fig:fig03} for the 
definition of initial states). In all cases, net charges are reported 
after subtracting the value at $t=0$; negative values for $\Delta q$ indicate
net flow of electron density
from the CN fragment to the metal center.
}                           
\label{fig:fig04}                                                  
\end{figure*}




\subsection{Dependence of Electron dynamics on Substitution Patterns}
\subsubsection{{\it m}-vs-{\it p} selectivity in electron dynamics of CN-C$_6$H$_4$-S-Au}
To induce CT selectivity in TDCI electron dynamics across 
linked-benzenes---an effect missing in cyanobenzene---we zero-in on the 
thiolate of benzonitrile bonded to an Au atom. To this end, we consider
both {\it m} and {\it p} terminated isomers. First of all, inspecting the 
electronic structure through the DOS near the valence energy,
reveals no apparent differences between the {\it p} and {\it m} 
isomers (see Fig.~\ref{fig:fig03}). Further, fragment-projected 
DOS reveals all the characteristic MOs localized on D and A 
fragments to have very similar energetics across 
both the isomers. By inspecting the three bonding-type MOs  
$\sigma$(CN-C), $\pi_o$(CN), $\pi_i$(CN)---subscripts $o$ and $i$ signify 
out-of-plane and in-plane w.r.t. the benzene plane---and their antibonding 
counterparts, we notice a flip in the phase of certain MOs of the {\it m}-isomer
compared to the {\it p} one. 

It is a well known fact based on HMO that {\it m} and {\it p} substitutions on a 
benzene ring lead to different phases for selected MOs. In the present work, we note based 
on HF calculations, that both $\pi_o$/$\pi_o^*$ MOs show a change in phase when 
the {\it m} link is replaced 
by a {\it p} one. On the other hand,  $\pi_i$/$\pi_i^*$ MOs conserve their phase
on both the isomers. Strikingly, for the $\sigma$-type MOs while the
low-energy bonding-type MO shows a phase-flip, the anti-bonding-type MO  does not 
suffer from phase changes. In the next section, we will illustrate how QI features,
hence CT selectivity, can be controlled in many-electron wavepacket 
dynamics via phase-flip effects in MOs. 

Having identified the MOs of interest, we consider as initial states 
for CT dynamics, CSFs formed by exciting an electron from an occupied MO to
an unoccupied one (see Fig.~\ref{fig:fig03}). It may be worthwhile to 
note that the energies of these 
CSFs are somewhat higher compared to the HOMO-to-LUMO excitation in these systems. 
All four CSFs
considered here show characteristics suitable to result in stable CT dynamics: 
i) excess electron density at the CN donor end
compared to the Au acceptor end at $t=0$, 
ii) availability of several unoccupied orbitals localized on Au essential for 
 trapping of electron density at the acceptor end for a few fs, and
iii) presence of MOs localized on the benzene fragment for conduction.

CT dynamics in {\it m} and {\it p} CN-C$_6$H$_4$-S-Au systems for three different initial
states are illustrated in Fig.\ref{fig:fig04}. Time evolution of the partial charge, $q$,
on the Au atom reveals selective electron-injection. Overall, one notes the
timescales for the events in all the cases to be about 3-4 fs.
As the most striking feature, we 
note CT mediated through the in-plane $\pi$ (CN) orbitals to be more efficient in the 
{\it m} isomer than in the {\it p} one.  However, the same process when 
mediated either through the 
$\pi_o$ or the $\sigma$ MO is more efficient in the {\it p} isomer. This
contrasting trend can be understood as follows: All three types of MOs are in resonance with the MOs localized at the Au acceptor terminal. In addition, the $\pi_o$ and $\sigma$ MOs are also in resonance with the MOs of same symmetry localized on the benzene 
fragment (the bridge) amounting to a resonant, through-bond CT process. 
This situation becomes more apparent through an inspection of the
MOs plotted in Fig.\ref{fig:fig03}, where one notes the densities of the $\pi_o$ MOs
to be predominantly localized at the donor end,
while small but non-vanishing MO densities localized through the bridge until the Au end.  
On the other hand, the $\pi_i$ MOs (bonding and antibonding) are strongly localized at the CN end without mixing of AOs from the benzene ring. Hence, CT mediated by
the $\pi_i$ MOs is a non-resonant tunneling process---the extent of which diminishes
with increase in the distance between the donor and acceptor terminals. Hence, in 
the {\it p}-isomer, one notes a drastic drop in the net electron transfer when
starting with the $\pi_i$-type initial state. 
The distance dependence of such a non-resonant tunneling process 
has been demonstrated experimentally\cite{batra2014molecular}
for two paracyclophane systems: one  where two benzene rings are connected 
at the {\it p}-ends by two methylene units ({\it i.e.} 22PCP) and another where the
 rings are connected  by four methylene units in a {\it p}-fashion ({\it i.e.} 44PCP). 
The inter-ring separation of the shorter and longer molecules 
are 3 and 4 \AA, respectively. The CT process in 44PCP
has been found 
to be 20 times slower than in 22PCP\cite{batra2014molecular}.
\begin{figure*}[hpt!] 
\centering          
\includegraphics[width=16cm, angle=0.0]{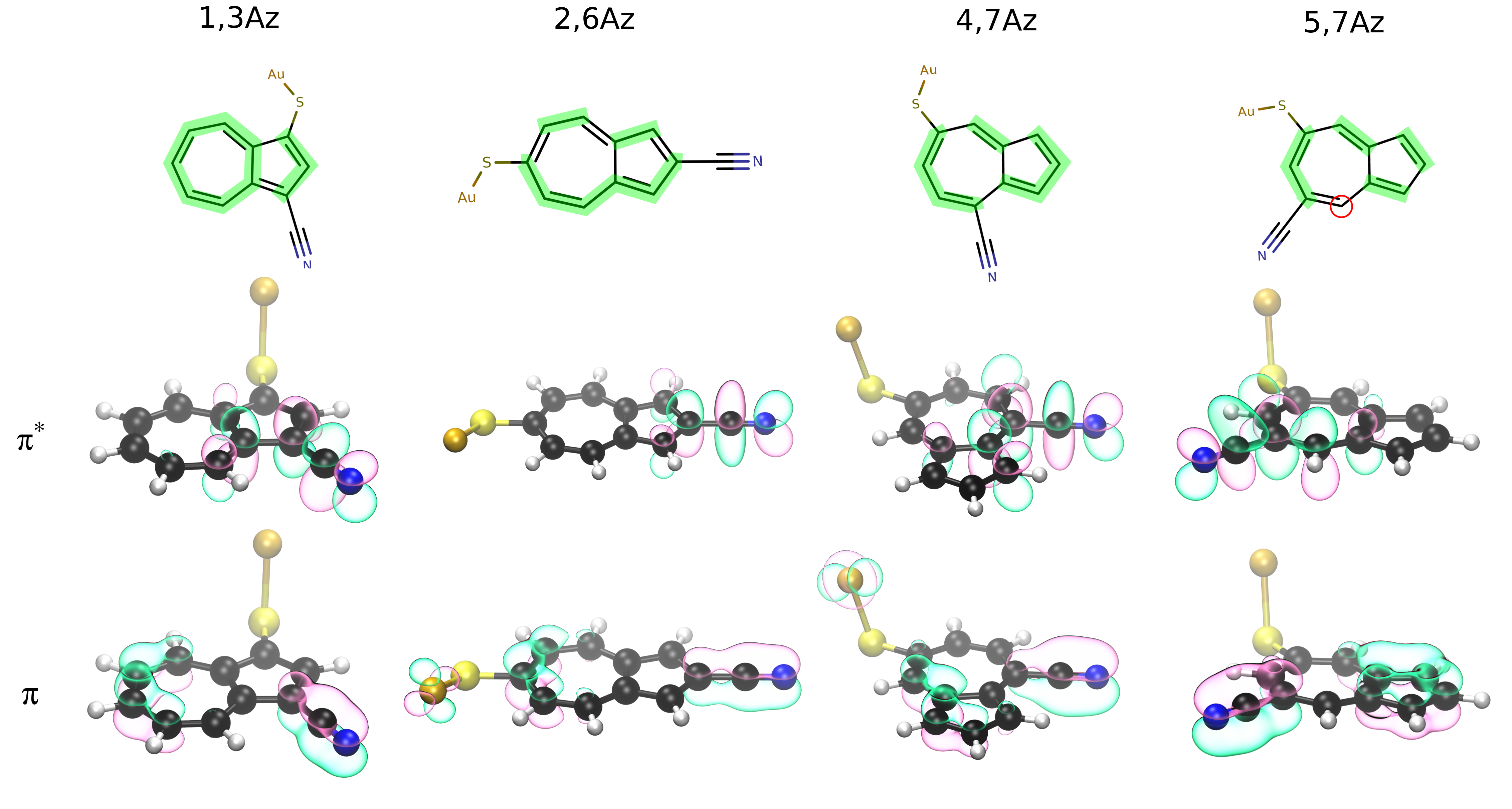}
\caption{
Isomers of gold-cyanoazulene thiolates with graphical scheme applied for the diagnosis of
destructive quantum interference. Also shown are the $\pi_o$ and $\pi^*_o$ MOs
localized on the CN fragment.
}
\label{fig:Azu1}
\end{figure*}

\subsubsection{Electron-transfer selectivity in gold-cyanoazulene thiolate isomers}
As a non-trivial case to study electron-transfer selectivity arising from different substitution
patterns, we zero-in on cyanoazulene thiolate linked to an Au atom (see Fig.~\ref{fig:Azu1}). 
Following conventional numbering of the C atoms, we denote the four isomers studied here:
$1,3{\rm Az}$, $2,6{\rm Az}$, $4,7{\rm Az}$, and $5,7{\rm Az}$, where ${\rm Az}$ is azulene. 
Conductivity of these isomers have been the subject of experimental and theoretical 
investigations\cite{xia2014breakdown,stadler2015comment,strange2015reply,schwarz2016charge,zhao2017destructive,tsuji2017frontier,wachter2017dynamics}. 

Xia {\it et al.}\cite{xia2014breakdown} have investigated 
the conductivity of the aforestated isomers using the STM-BJ 
technique and NEGF-DFT calculations. This study
depended on Az molecules connected non-covalently to gold junctions
through dimethylthiochroman anchor resulting in small numerical values of $G$
---in the range $8\times10^{-5} G_0$ to $32\times10^{-5} G_0$---when averaged over measurements; 
the overall trend in conductivity has been concluded as 
${2,6{\rm Az}} \approx {1,3{\rm Az}} > {4,7{\rm Az}} > {5,7{\rm Az}}$. In the same work, 
the authors reported $G$ from NEGF-DFT calculations by setting the Fermi energy, 
$E_{\rm F}$, to -1.5 eV and
noted the trend ${1,3{\rm Az}} > {2,6{\rm Az}} \approx {4,7{\rm Az}} > {5,7{\rm Az}}$ agreeing 
semi-qualitatively with experimental trends. However, $G$ 
of $1,3{\rm Az}$ has been noted to drop considerably in 
an NEGF-DFT calculation with $E_{\rm F}$=0 eV.
Later, Stadler\cite{stadler2015comment} had reiterated the criterion for
finite conductance and vanishing
destructive QI at $E_{\rm F}=0$ eV to be: all the AOs of the molecular topology
should either lie on a continuous path connecting the terminals or lie on a closed loop.
As seen in Fig.~\ref{fig:Azu1}, ${1,3{\rm Az}}$ does not feature unpaired atomic centers
indicating the absence of destructive quantum interference.
In a separate study, Strange {\it et al.}\cite{strange2015reply}
have performed NEGF-DFT calculations for three more isomers, 
${1,4{\rm Az}}$, ${1,6{\rm Az}}$, and ${1,8{\rm Az}}$---all of them 
satisfying the connectivity-based conditions for finite conductance---to 
exhibit vanishing transmission at $E_{\rm F}$=0 eV.
This observation strengthened the notion that in these isomers, and in ${1,3{\rm Az}}$, 
factors other than interference effects are responsible for
a drop in the magnitude of $G$ at $E_{\rm F}$=0 eV.
In the case of $1,3{\rm Az}$, further clarity emerged from the combined 
experimental and theoretical work of
Schwarz {\it et al.}\cite{schwarz2016charge} who studied derivatives of Az that are covalently
bonded to the terminals. Compared to the earlier values for a dimethylthiochroman 
anchor\cite{xia2014breakdown}, ${2,6{\rm Az}}$ and
${4,7{\rm Az}}$ exhibited better conductance with
$G=5\times10^{-2} G_0$ while $1,3{\rm Az}$ showed $G=1.6 \times10^{-4} G_0$. 
Overall, the experimental trend in $G$ for the molecules 
displayed in  Fig.~\ref{fig:Azu1},
at low $T$ and vanishing electron-phonon coupling, follows
${2,6{\rm Az}} \approx {4,7{\rm Az}} > {1,3{\rm Az}}$.





\begin{figure}[hpt!] 
\centering                                                                 
\includegraphics[width=8.0cm, angle=0.0]{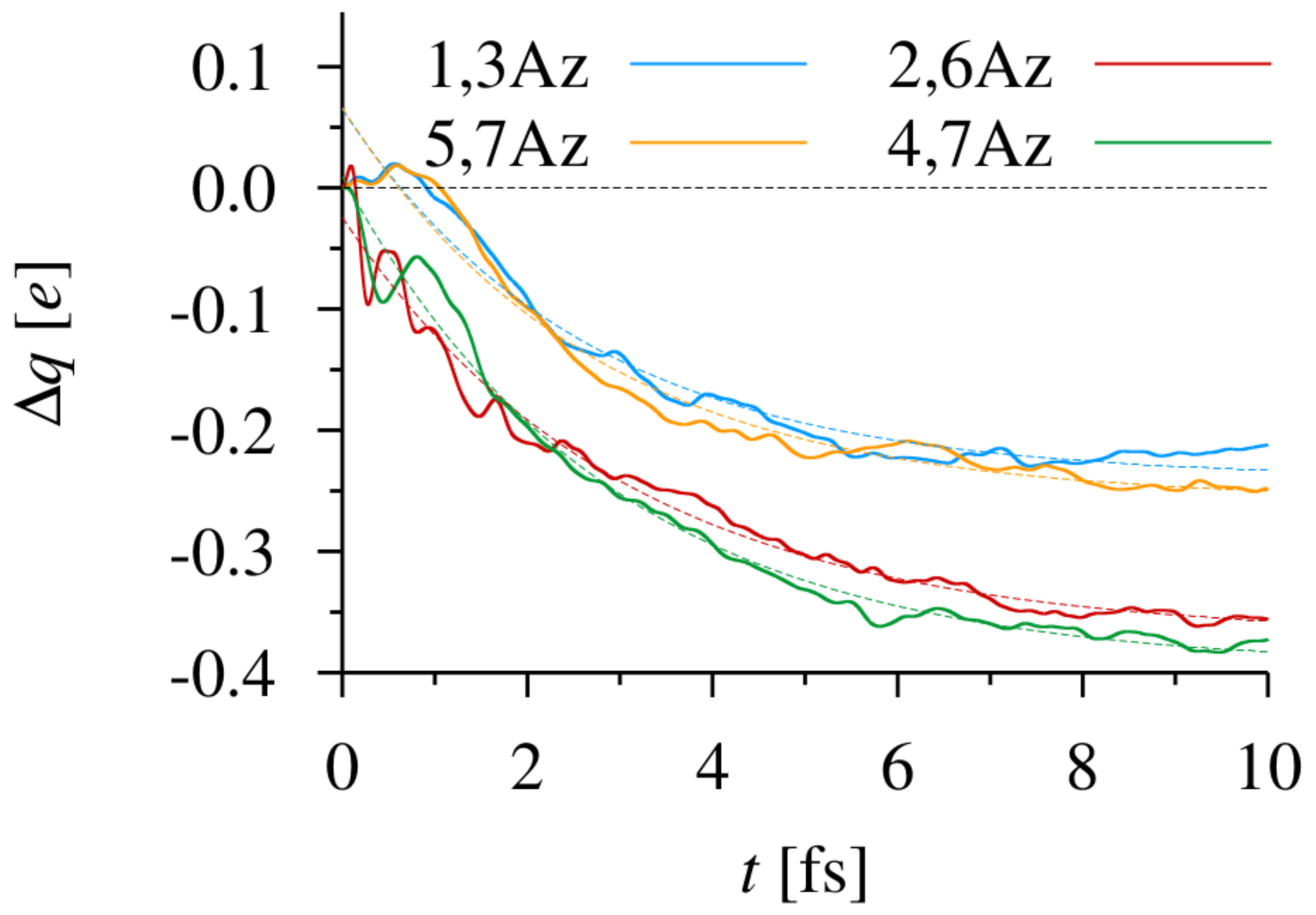}
\caption{
Time evolution of the partial charge, $q$, on the Au atom 
in CN-C$_{10}$H$_6$-S-Au isomers; 
net charge ($\Delta q$) is reported 
after subtracting the value at $t=0$; negative values indicate
flow of electron density
from the CN fragment to the metal center.
In all cases, the 
initial state, $|\Psi(0)\rangle$ is created by exciting an electron from the $\pi_o ({\rm CN})$ MO
to the $\pi_o^* ({\rm CN})$ MO. 
}                           
\label{fig:Azulene}                                                  
\end{figure}

TDCISD CT dynamics of gold cyanoazulene thiolate isomers are shown in 
Fig.~\ref{fig:Azulene}. In all cases, the initial state
is a CSF corresponding to $\pi_o \rightarrow \pi_o^*$ excitation; the MOs involved
are on display in Fig.~\ref{fig:Azu1}. During the first 10 fs of the dynamics, the 
time-evolved partial charge on the Au atom
follows the first-order-type relation  $\Delta q(t) = \Delta q_0 \exp(-t/\tau) + \Delta q_\infty $. 
CT parameters estimated via a least squares fitting 
of $\Delta q (t)$ are collected in
Table~\ref{tab:Azu2}; inspection of these results reveal the timescale for electron migration to
lie in the narrow range 2.6--3.0 fs. The most striking qualitative trend as seen in 
Fig.~\ref{fig:Azu1} and Table~\ref{tab:Azu2} is that both $2,6{\rm Az}$ and $4,7{\rm Az}$ 
isomers show larger gain of elecron density at the Au terminal compared to 
$1,3{\rm Az}$ and $5,7{\rm Az}$ isomers. The trend in $\Delta q_\infty$ can be seen to follow 
the aforementioned trend noted in experimentally determined $G$:
${2,6{\rm Az}} \approx {4,7{\rm Az}} > {1,3{\rm Az}} > {5,7{\rm Az}}$.
As noted in a previous TDCI study\cite{ramakrishnan2013electron} 
of Li-terminated cyano-alkenes and -alkynes, a drop in $\Delta q_\infty$ is a 
consequence of either all the MOs involved being fully delocalized from the donor-terminal to the
acceptor one, or even when the $\pi$-type-MOs are sufficiently localized on the donor-terminal, 
truncation in the MO network of suitable symmetry from donor-to-bridge-to-acceptor 
making CT to proceed only via distance-dependent tunneling. The later mechanism which is on 
action in {\it m}-linked benzene, as discussed above, also controls the dynamics of ${1,3{\rm Az}}$ and $5,7{\rm Az}$.

\begin{table}[htp]
    \centering
    \caption{Charge-transfer parameters for gold-cyanoazulene thiolate isomers 
     obtained by fitting the time-evolution of the net-electron gain, $\Delta q$,
     at the Au terminal
     to $\Delta q(t) = \Delta q_0 \exp(-t/\tau) + \Delta q_\infty $.
     Net charges are reported in $e$ and the timescale is in fs.
     } 
    \begin{tabular}{l l l l}
    \hline
    Molecule~~~~~~~~~~&$\Delta q_0$~~~~~~~~~~& $\Delta q_\infty$~~~~~~~~~~&$\tau$\\
    \hline
    $1,3{\rm Az}$     &~0.06                 &-0.24                       & 2.62 \\
    $2,6{\rm Az}$     &-0.02                 &-0.37                       & 3.03 \\
    $4,7{\rm Az}$     &~0.01                 &-0.40                       & 2.88 \\
    $5,7{\rm Az}$     &~0.07                 &-0.26                       & 2.68 \\
    \hline
    \end{tabular}
    \label{tab:Azu2}
\end{table}

\subsection{Quantum interference via linear superposition of many-body wavefunctions}

\begin{figure*}[hpt!] 
\centering 
\includegraphics[width=16.0cm, angle=0.0]{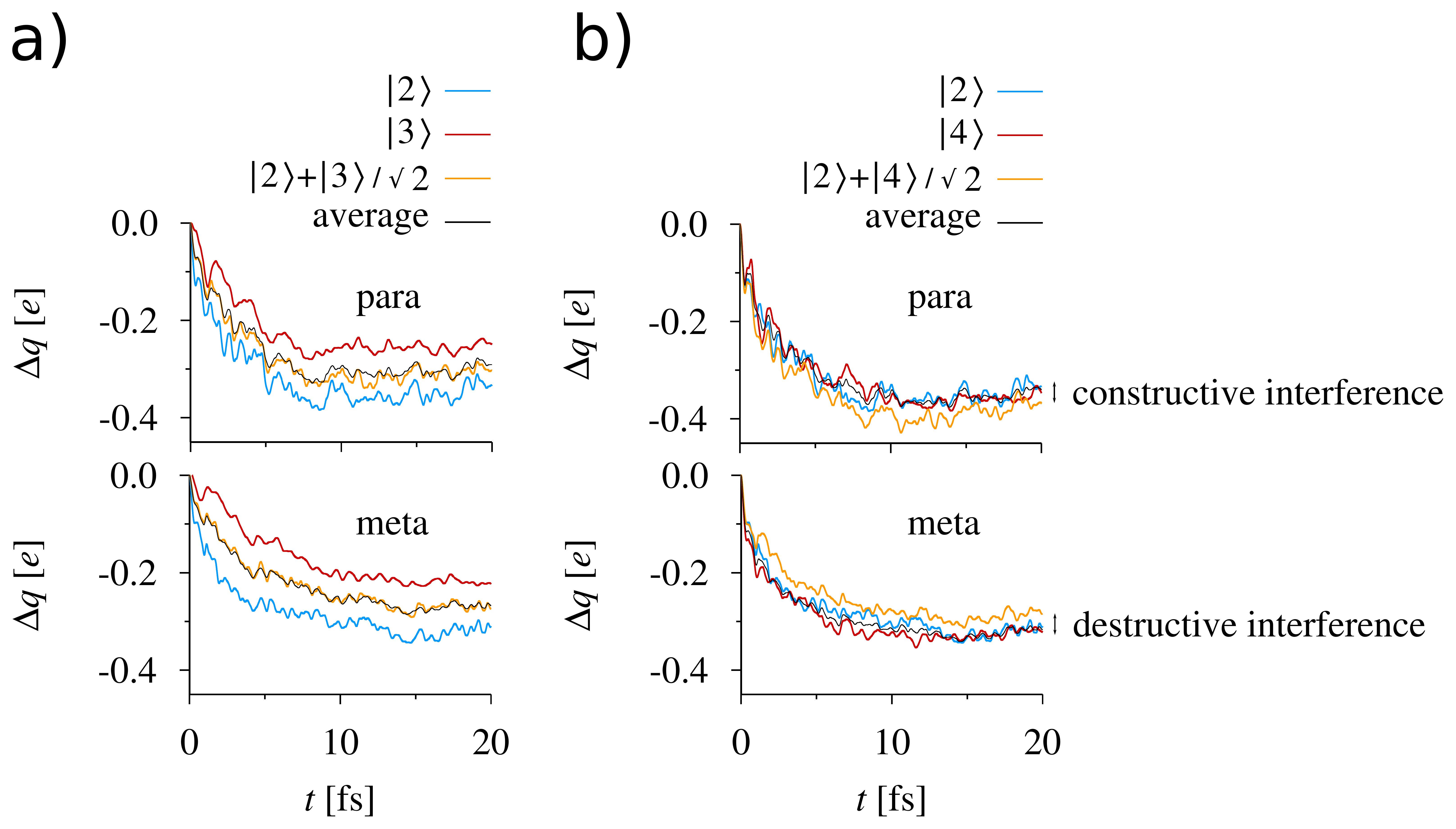}                                       
\caption{                
Time evolution of net-electron gain, 
$\Delta q$, at the Au atom 
in {\it m}- and {\it p}-linked CN-C$_6$H$_4$-S-Au.
Results are shown for two choices of
initial states that are coherent superpositions:
a) Dynamics of  $\left[ |2\rangle+|3\rangle \right]/2$ coincides with the average of the values from the individual states $|2\rangle$ and $|3\rangle$ indicating negligible interference effects. 
b) $\Delta q(t)$ of the superposed state showing enhanced 
electron migration for the {\it p} isomer and a suppressed migration 
for the {\it m} isomer compared to the average of the values from the 
individual states $|2\rangle$ and $|4\rangle$. 
}                                                                          
\label{fig:fig05}
\end{figure*}
Tsuji {\it et al.}\cite{tsuji2017frontier,tsuji2018quantum} have shown and discussed with
great clarity that in conductivity calculations based on the HMO model Hamiltonian, QI can be
predicted for a pair of atoms $\mu$ and $\nu$ using the zeroth-order Green's function 
\begin{eqnarray} 
  G_{\mu\nu}^0 (E_{\rm F}) = \sum_{k\,{\rm \in \, MOs}}
\frac{C_{\mu k}C_{\nu k}^*}{E_{\rm F} - E_k + i \eta},
\end{eqnarray} 
where $E_{\rm F}$ is the Fermi energy, $E_k$ is the energy of the $k$-th MO and 
$\eta$ is a small positive number. The molecular junction features destructive QI 
when $G_{\mu\nu}^0 (E_{\rm F})$ vanishes for $E_{\rm F}=0$.
For any pair of MOs (say $i$-th MO is $\pi$ and $j$-th is $\pi^*$),
with an assumption that $E_{\rm F}$ lies between the energy levels of
the frontier MOs (FMOs), {\it i.e.}, $E_i<0$ and $E_j>0$, 
$G_{\mu\nu}^0 (E_{\rm F})$ diminishes when
sgn({$C_{\mu i}C_{\nu i}^*$}) = sgn({$C_{\mu j}C_{\nu j}^*$}), where sgn() is the 
sign function. On the other hand, for the situation 
sgn({$C_{\mu i}C_{\nu i}^*$}) = -sgn({$C_{\mu j}C_{\nu j}^*$}), 
the contributions to the Green's function from FMOs $i$ and $j$ 
add up amounting to better transmission. Following this argument, 
it is rather straightforward to see how the $\it m$-vs.-${\it p}$ selectivity
can be interpreted as arising from the phase-flip of one of the FMOs in the 
{\it m} isomer compared to the {\it p} counterpart. 
Koga {\it et al.}\cite{koga2012orbital} have shown using DFT-NEGF calculations and symmetry 
arguments how the aforestated selectivity is preserved when coupling the benzene
bridge to $\pi$-acceptor anchor groups.

To realize QI in real-time electron dynamics simulations, we consider 
initial states that are linear superpositions of CSFs. When following the time evolution
of any quantum mechanical observable, interference effects arise from the 
off-diagonal (or coherence) terms in the expectation value. If the initial state is given by the symmetric linear combination 
$|\Psi(0)\rangle=\left[ |1\rangle+|2\rangle\right]/\sqrt{2}$, 
the expectation value of an observable $O$ is given by 
$<\hat{O}>=<\hat{O}>_{\rm ave.} + <\hat{O}>_{\rm int.}$, where the 
first term denotes averaging over both states, 
$<\hat{O}>_{\rm ave.}=\left[ O_{11} + O_{22} \right]/2$,
while the second term arises due to interference,
$<\hat{O}>_{\rm int.}=\left[ O_{12} + O_{21} \right]/2={\mathcal Re} \left[O_{12}\right]$.

To form initial states that are linear superpositions, we  consider 
three CSFs (Fig.\ref{fig:fig03}): 
$ | 2 \rangle$ (corresponding to  $\pi_o\rightarrow\pi_o^*$ excitation),
$ | 3 \rangle$ (corresponding to $\sigma\rightarrow\sigma^*$ excitation) and 
$ | 4 \rangle$ (corresponding to $\pi_o \rightarrow\sigma^*$ excitation).
Fig.\ref{fig:fig05} features time-dependent partial charge on the Au acceptor terminal
for two choices of initial states: 
$| \Psi(0) \rangle = [| 2 \rangle + | 3 \rangle]/\sqrt{2}$ and 
$| \Psi(0) \rangle = [| 2 \rangle + | 4 \rangle]/\sqrt{2}$.
Starting with the first option, we see that both in the {\it m} and {\it p}
isomers, CT dynamics follow the average of the dynamics exhibited
separately by states $| 2 \rangle$ and $| 3 \rangle$ indicating vanishing
contributions from the interference terms. This can be understood taking into account
the Slater--Condon rules to evaluate matrix elements for a one-electron
operator, $\mathcal{O}_1$\cite{szabo1996modern}. Accordingly, at $t=0$ fs, direct terms
contributing to $q(t)$ can be determined as 
$\langle  \Psi_a^r | \mathcal{O}_1 | \Psi_a^r \rangle = 
\sum_c^N \langle c |\mathcal{O}_1 | c \rangle - \langle a | \mathcal{O}_1 | a \rangle + \langle r | \mathcal{O}_1 | r \rangle $. As for the coherence terms, 
contributions arise from the 
matrix elements of the form $\langle  \Psi_a^r | \mathcal{O}_1 | \Psi_b^s \rangle$.
This matrix element vanishes according to the Slater--Condon rules\cite{szabo1996modern} when $a \ne b;\,r \ne s$. 
Non-vanishing contributions to interference terms arise  only when 
$a = b;\, r \ne s;\,\langle  \Psi_a^r | \mathcal{O}_1 | \Psi_a^s \rangle = \langle r | \mathcal{O}_1| s \rangle$ or 
$a \ne b;\, r = s; \, \langle  \Psi_a^r | \mathcal{O}_1 | \Psi_b^r \rangle = -\langle b | \mathcal{O}_1| a \rangle$. 
While inspecting the time-evolution of $\Delta q$ at the Au terminal, starting
with the superposed initial state $| \Psi(0) \rangle = [| 2 \rangle + | 4 \rangle]/\sqrt{2}$,
we note in the case of the {\it p} isomer, superior net CT
compared to the average of the dynamics exhibited
separately by $| 2 \rangle$ and $| 4 \rangle$. In contrast, 
for the {\it m} isomer, $\Delta q$ drops noticeably compared to the average dynamics.
This trend suggests that QI contributions in real-time dynamics with many-electron
wavefunctions can feature both constructive and destructive QI effects depending
on the sign of the coherence contributions to the time-dependent expectation values.

\section{Conclusions}
In summary, following the real-time, many-body method 
TDCI offers an exact, all-electron and 
many-body picture of ultrafast electron dynamics in 
D-B-A systems.
The dynamics is sensitive to the choice of the initial 
state. In cyanobenzene, when starting with a state 
created by $\pi\rightarrow\pi^*$ excitation, electron density
is injected into the benzene fragment within the first 5 fs. 
During the initial part of time-evolution, the net CT is 
maximal at the $o$-position. By $t=4$ fs and $t=5$ fs, electron density
reaches $m$- and $p$-sites, respectively, with essentially no preference for one site
over the other beyond that demanded by distance.  
The CT process is oscillatory with 
very short timescales typical of wavepacket evolution in finite systems.
Attaching the benzene molecule to an acceptor terminal linked at 
{\it m} or {\it p} position stabilizes the CT process 
and delays wavepacket revival. 
Dynamics involving $\pi_o$ or $\sigma$ MOs, localized on the CN fragment, show enhanced CT in the 
{\it p} isomer compared to the {\it m} one. On the other hand, dynamics along 
the $\pi_i$ channel shows counter-intuitive selectivity, where the CT in the {\it m} isomer
is more efficient than the {\it p} isomer due to non-resonant tunneling that drops rapidly with increase in the distance between D and A terminals. 
TDCI dynamics distinguish $1,3{\rm Az}$, $2,6{\rm Az}$, $4,7{\rm Az}$, and $5,7{\rm Az}$
isomeric bridges into two classes: 
(i) Those with conjugated $\pi$-type MO network on donor, bridge and acceptor sites;
(ii) Those where the $\pi$-type MO network is disrupted.  
$2,6{\rm Az}$ and  $4,7{\rm Az}$ isomers belong to the former class permitting 
large net CT while $1,3{\rm Az}$ and $5,7{\rm Az}$ isomers belonging to the latter case 
exhibit small net CT arising from non-resonant tunneling. 
This observation clarifies the poor conductivity of the $1,3{\rm Az}$ isomer, 
which has been noted as a pathological case in transport
studies because of the sensitivity of its $G$ with the change in the anchor group
coupling azulene to metal terminals\cite{xia2014breakdown,stadler2015comment,strange2015reply,schwarz2016charge}.

Compared to the Green's function and density matrix formalisms, where QI
features appear due to cancellation of phases in different paths, in
the many body approach presented here, we see QI appearing from
 many-electron wavefunctions that are 
spread over all possible paths. 
The TDCI formalism can be adapted to model finite-bias conductance
to study metal-molecule-metal junctions. 
Such a formalism based on localized density constraints to create a chemical 
potential bias has been developed in the framework of RT-DFT 
and has been shown to give $I$-$V$ curves of a molecular wire
in agreement with Green's function calculations\cite{cheng2006simulating}. 
This procedure when used with a many-body formalism like TDCI, 
besides providing quantitative 
state-selective details that are not accessible in DFT-based 
Green's function calculations, can also address ambiguities 
that arise in single-determinant electron dynamics\cite{ramakrishnan2012control}. 
Modifications can also be made
to the choice of the junction contacts by replacing 
the thiolate group with an amine group that has shown to result in more 
reproducible conductance\cite{venkataraman2006single}.

\section{Acknowledgments}
The author gratefully thanks Ravi Venkatramani and his group members for 
countless discussions on charge-transfer phenomena, Reviewer-2 for
thought-provoking comments, and Salini Senthil for assistance with the
manuscript revision. 
This project was funded by intramural funds at TIFR Hyderabad from 
the Department of Atomic Energy (DAE).
All calculations have been performed using the Helios computer cluster, 
which is an integral part of the MolDis Big Data facility, 
TIFR Hyderabad (https://moldis.tifrh.res.in/).
\bibliography{lit} 
\end{document}